\documentstyle[12pt,aasms4,epsfig]{article}

\def\beq{\begin{equation}}
\def\eeq{\end{equation}}
\def\bey{\begin{eqnarray}}
\def\eey{\end{eqnarray}}

\def\rel{{\rm rel}}
\def\msun{\;{\rm M}_\odot}
\def\h{{\rm h}}
\def\m{{\rm m}}

\def\thalf{t_{1/2}}
\def\tp{t_{\rm 0}}

\def\Ap{A_{\rm max}}

\def\te{t_{\rm E}}
\def\thetae{\theta_{\rm E}}

\def\base{{\rm base}}

\def\peak{{\rm peak}}

\def\spose#1{\hbox to 0pt{#1\hss}}
\def\lta{\mathrel{\spose{\lower 3pt\hbox{$\sim$}}
    \raise 2.0pt\hbox{$<$}}}
\def\gta{\mathrel{\spose{\lower 3pt\hbox{$\sim$}}
    \raise 2.0pt\hbox{$>$}}}

\input epsf

\begin{document}

\title
        {A Short Timescale
Candidate Microlensing Event in the POINT-AGAPE Pixel Lensing
        Survey of M31}

\author{M.~Auri{\`e}re$^1$, P.~Baillon$^2$, A.~Bouquet$^3$, B.J.~Carr$^4$, 
M.~Cr{\'e}z{\'e}$^{3,5}$, N.W.~Evans$^6$, Y.~Giraud-H{\'e}raud$^3$, 
A.~Gould$^{3,7}$, 
P.~Hewett$^8$, J. Kaplan$^3$, E.~Kerins$^6$, 
E.~Lastennet$^{4,9}$, Y.~Le~Du$^6$, A.-L.~Melchior$^{4,10}$,
S.~Paulin-Henriksson$^3$, S.J.~Smartt$^8$ and D.~Valls-Gabaud$^{11}$ }
\centerline{(The POINT--AGAPE Collaboration)}

\noindent
$^1$Observatoire Midi-Pyr{\'e}n{\'e}es, 57 Avenue d'Azereix, BP 826,
    65008 Tarbes Cedex, France \\
$^2$CERN, 1211 Gen{\`e}ve, Switzerland \\
$^3$Laboratoire de Physique Corpusculaire et Cosmologie, 
Coll{\`e}ge de France, 11 Place Marcelin Berthelot, F-75231 Paris, France\\
$^4$Astronomy Unit, School of Mathematical Sciences, Queen Mary
    \& Westfield College, Mile End Road, London E1 4NS, UK\\
$^{5}$Universit\'e Bretagne-Sud, campus de Tohannic, BP 573, F-56017 
    Vannes Cedex, France\\
$^6$Theoretical Physics, 1 Keble Road, Oxford OX1 3NP, UK\\
$^7$Department of Astronomy, Ohio State University, 140 West 18th Avenue, Columbus, OH 43210\\
$^8$Institute of Astronomy, Madingley Road, Cambridge CB3 0HA, UK\\
$^9$Depto. de Astronomia, UFRJ, Ladeira do Pedro Ant\^onio 43,
20080-090 Rio de Janeiro RJ, Brazil \\
$^{10}$DEMIRM UMR~8540, Observatoire de Paris, 61 Avenue
    Denfert-Rochereau, F-75014 Paris, France\\
$^{11}$Laboratoire d'Astrophysique UMR~CNRS~5572, Observatoire
    Midi-Pyr{\'e}n{\'e}es, 14 Avenue Edouard Belin, F-31400 Toulouse,
    France \\
\label{firstpage}

\begin{abstract}
We report the discovery of a short-duration microlensing candidate in
the northern field of the POINT-AGAPE pixel lensing survey towards
M31.  The full-width half-maximum timescale is very short, $\thalf =
1.8\,$days.  Almost certainly, the source star has been identified on
{\it Hubble Space Telescope} archival images, allowing us to infer an
Einstein crossing time of $\te=10.4\,$days, a maximum magnification of
$\Ap\sim 18$, and a lens-source proper motion
$\mu_\rel>0.3\,\mu$as/day.  The event lies projected at $8'$ from the
center of M31, which is beyond the bulk of the stellar lens
population.  The lens is likely to reside in one of three
locations. It may be a star in the M31 disk, or a massive compact halo
object (Macho) in either M31 or the Milky Way.  The most probable mass
is $0.06 \msun$ for an M31 Macho, $0.02 \msun$ for a Milky Way Macho
and $0.2\msun$ for an M31 stellar lens.  Whilst the stellar
interpretation is plausible, the Macho interpretation is the most
probable for halo fractions above $20 \%$.
\end{abstract}


\section{Introduction}

Following the suggestion of Paczy\'nski (1986), several groups
searched for dark matter in the form of massive compact halo objects
(Machos), using gravitational microlensing of background stars in the
Magellanic Clouds (see Alcock et al.\ 2000; Lasserre et al.\ 2000; and
references therein). After monitoring about 10$^{7}$ stars for several
years, the results are consistent with a Macho halo mass fraction of
$20\%$, though with considerable uncertainty.  

The possibility of detecting microlensing events in M31 was
independently suggested by Crotts (1992) and Baillon et al.\
(1993).  The advantage of targeting a large external galaxy is that
the number of stars that can act as possible sources is
enormous. Moreover, the high inclination of M31's disk causes an
asymmetry in the observed rate of microlensing by lenses in a
spheroidal halo. Although this gives an unambiguous signature of the
halo lenses (\cite{cro92}), the difficulty is that the sources are
resolved only while they are lensed (and then only if the
magnification is substantial).  Nonetheless, pilot campaigns by the
Columbia-VATT group (\cite{cro97}) and by the AGAPE collaboration
(\cite{ans97}; \cite{ans99}) established the feasibility of such
observations and identified some candidate microlensing events. In
particular, the AGAPE collaboration (\cite{ans99}) reported a short
duration candidate, AGAPE Z1, in the bulge of M31.

The POINT-AGAPE collaboration\footnote{ see {\tt
http://www.point-agape.org}} employs the Wide Field Camera (WFC) on
the 2.5 m Isaac Newton Telescope (INT) to carry out a pixel-lensing
survey towards M31, monitoring two fields of 0.3 deg$^2$ each, located
North and South of the M31 center.  The survey has the potential to
map the global distribution of the microlensing events in M31 and to
determine any large-scale gradient.  POINT-AGAPE is a long-term
program, as at least three years of data will be required to make a
convincing identification of a gradient (Kerins et al.\ 2001). Also,
it is usually necessary with pixel lensing to establish a long
baseline to distinguish microlensing events from variable stars.  In
this {\it Letter}, we report on a particularly interesting and
convincing early candidate, PA-99-N1, so named because it is the first
event to be announced by POINT-AGAPE that peaks in the northern field in 1999.

\section{Observations and Data Analysis} 

We restrict analysis to a ($11'\times 22'$) field centered
$3'$ west and $12'$ north of the center of M31.  The
observations  
are spread over 36 epochs between August and December 1999. The
exposures are in two bands: 36 epochs in Sloan $r^\prime$ and 26
epochs in Sloan $g^\prime$. The exposure time is typically between 5
and 10 minutes per night. The sampling averages to one epoch every
three nights, though the observations are strongly clustered because
the WFC is not always mounted. The data reduction is described in
detail elsewhere (\cite{ans97}; \cite{naples98}; \cite{ledu00}).

After bias subtraction and flat-fielding, each image is geometrically
and photometrically aligned relative to a reference image (14 August
1999) which was chosen because it has a long exposure time, typical
\mbox{seeing ($1.\hskip-2pt ''6$)} and little contamination from the
Moon.  The lightcurves are computed by summing the flux in 7-pixel
($2.\hskip-2pt ''3$) square ``superpixels'' and then removing the
correlation with seeing variation.  This size is set by the worst
seeing $\sim 2.\hskip-2pt ''1$.

We use a simple set of candidate selection criteria designed to
isolate high signal-to-noise events.  Detection of events is made in
the $r'$ band, which has better sampling and lower sky background
variability.  For an event to be detected, it must induce one and only
one significant bump on the lightcurve. A bump is defined by at least
4 consecutive data points rising above the background by a minimum of
$4\sigma$. Its significance is quantified by the probability $P$ of
obtaining data points, each of which has at least the signal to noise
ratio of the measured points (evaluated assuming Gaussian errors). To
pass as a microlensing candidate, we demand the lightcurve has only
one bump with $-\mbox{ln} P > 100$ and no other bump with $-\mbox{ln}
P > 20$. A similar selection procedure was used by the AGAPE Pic du
Midi program (Ansari et al. 1997).

The background is the minimum of the running average of 7
consecutive data points.  Flagged events are fitted to a high
magnification degenerate microlensing curve (Gould 1996)
simultaneously in the \mbox{$r'$} and \mbox{$g'$} bands,
whose 6 parameters are
($\thalf,t_0,F_{\base,r'},
F_{\base,g'},\Delta F_{\peak,r'},\Delta F_{\peak,g'}$).
Here, $\thalf$ is the full-width at half-maximum, $t_0$ is the time
of the peak, $F_\base$ is the baseline flux and $\Delta F_\peak$ is
the flux difference between baseline and maximum.  We require 
a $\chi^2$ per degree of freedom be less than 3 for this fit. 
We also require at least minimal detection in $g'$, i.e.
$\Delta F_{\peak,g'}>1\,\rm ADU\,s^{-1}$.
To insure a high
probability of detecting microlensing rather than other forms of
stellar variability, given the clustered sampling and relatively short
duration of the observations, we demand $\thalf<8\,$days and a peak
flux $\Delta F_{\peak,r'}>10\,\rm ADU\,s^{-1}$, corresponding to 
$R_{\rm peak}<21.5$.  
We find one candidate which has $\thalf\sim 1.8\,$day
and $\Delta F_{\peak,r'}\sim 17\,\rm ADU\,s^{-1}$.  

\section{The Microlensing Candidate}

Figure \ref{fig:lightcurves} shows the lightcurves in $r'$ and $g'$ of
this candidate together with the non-degenerate fit derived below.
The $g'$ data with comparatively large error bars were taken on nights
with high Moon background.  Using the Aladin Sky Atlas\footnote{ see
{\tt http://aladin.u-strasbg.fr/aladin.gml}}, we find that PA-99-N1
has J2000 position: \mbox{$\alpha = 00^\h42^\m51\fs 42$},
\mbox{$\delta = +41^\circ 23' 53\farcs 7$}.  That is, it lies
projected on the near disk, $7^{\prime} 52^{\prime\prime}$ from the
center of M31.

There are some straightforward tests to see if PA-99-N1 is compatible
with microlensing. First, there are no comparable ``bumps'' in the
remainder of the lightcurve shown in Figure \ref{fig:lightcurves} as
might be expected for many classes of variable stars. (There is,
however, a much smaller but somewhat disturbing bump in December, near
day 130.  We investigate this bump in \S\ 4.1.)\ \ There also are no
comparable bumps at this field position in our data taken in 1998 and
1999 with the 1.3m McGraw-Hill telescope at the MDM
observatory. Second, microlensing events are achromatic, so the ratio
of flux {\em change}\/ in different bands should be constant in time
(e.g., Ansari et al.\ 1997).  This requires
\begin{equation}
{{\Delta F_{g^\prime}(t)} \over {\Delta F_{r^\prime}(t)}} = 
{{F_{g^\prime}(t) - F_{\base,g^{\prime}}} \over 
{F_{r^\prime}(t) - F_{\base,r^{\prime}}}} = {\rm constant},
\end{equation}
which does indeed hold for PA-99-N1 (see Fig.\ \ref{fig:lightcurves}c).

Using DAOPHOT (\cite{stetson}) on the images taken nearest maximum
magnification, we find $R = 20.80 \pm 0.13$ and $V = 22.00 \pm 0.17$,
i.e., $V-R=1.20\pm 0.22$ and $M_V=-2.8\pm 0.3$, assuming
$(m-M)_{0,M31}=24.43$ and estimated total extinction
$A_V=0.4\pm 0.2$.  The transformation from
instrumental $(r',g')$ to Johnson $(V,R)$ is based on 31 standard
stars that lie in the same field as PA-99-N1 (\cite{magone};
\cite{magtwo}; \cite{haiman}).

The PA-99-N1 position lies within a series of five {\it Hubble Space
Telescope (HST)} WFPC2 archival images taken in July 1996, three with
F814W ($\sim I$) and two with F606W ($\sim V$).  We use the relations
of Zheng et al.\ (2001) to transform from these filters to
Johnson-Cousins $V$ and $I$.  Since the event is very red,
$V-R\sim 1.2$, the source must lie either high up on the giant branch
where essentially all stars are resolved by {\it HST}, or must be on
the main-sequence and so magnified by $\Ap\ga 10^4$ (see Fig.\ \ref{hst1}).
The latter possibility is extremely unlikely a priori.  Hence, a firm
prediction of the microlensing interpretation is that there should be
a resolved star in the {\it HST} image with the same color and same
position as the event.

To measure the spatial position of the event within the INT image we
take the difference of the image at maximum and another image at
baseline that has similar seeing.  The difference image shows a
well-defined and isolated difference star whose position we measure
using IRAF IMEXAMIN.  Variation of the IMEXAMIN parameters changes the
result by $\sim 0.2$ pixels ($0.\hskip-2pt ''07$), which we adopt as
the error in the measurement.  The spatial transformation between the
INT and {\it HST} fields is derived by comparing the positions of the
eight stars that are both resolved in the INT image and unsaturated in
the {\it HST} image.  The uncertainty in this transformation is much
smaller than that of the INT position measurement.
Figure~\ref{hststar} shows part of the {\it HST} field with the $1
\sigma$ and $3\sigma$ positional error circle 
centered on the event. Close to this
circle there is a resolved star for which we find $V = 24.51 \pm 0.12$
and $I = 22.41 \pm 0.10$, implying $V-I = 2.10 \pm 0.16$.  There are
no other resolved stars within the $3\,\sigma$ circle.  For typical
stellar populations this $V-I$ is compatible with the $V-R=1.20\pm
0.22$ measured for the event (Demarque et al.\ 1996; Yi, Demarque, \&
Oemler 1996).  The prior probability to find such a star so close to
the predicted position is only 3\%.  We conclude that the resolved HST
object is almost certainly the source star of the event.

\section{Lightcurve Interpretation}

\subsection{PA-99-N1 as a Variable Star}

	Before analyzing PA-99-N1 as a microlensing event, we need to 
ask whether it is consistent with an interpretation as a variable star?  The
color and magnitude are compatible with a M0 Mira at maximum.
However, such Miras have periods greater than 200 days and their flux
at maximum does not vary as rapidly as the event (Allen 1999).
PA-99-N1 is unlikely to be a dwarf nova. It is too bright at maximum
to be in M31. If it is a Galactic dwarf nova, it must be either
unusually faint or improbably far from us ($> 10$ kpc). 
It is also
unlikely to be a nova. The brightness decrease of PA-99-N1 just after
maximum is very rapid with a rate of decline of about 0.7
magnitudes/day. Capaccioli et al.\ (1989) studied the relation between
the rate of decline and the magnitude at maximum for novae in M31 and
found that the brighter the nova, the faster the decrease.  The
decline of PA-99-N1 would indicate a nova as bright as $V=16$, which
would imply $A_V=6$, in stark contrast to the small extinction seen in
Figure \ref{hst1}.  To conclude, there is no type of stellar
variability known to us that could generate the lightcurve of
PA-99-N1.

	However, if the December bump near day 130 in Figure
\ref{fig:lightcurves} were due to the event source, one would
nevertheless have to conclude that the source was probably a variable.
We therefore investigate this bump closely.  For each of the nine
December images, we find a corresponding image at baseline with
comparable seeing, and subtract the two.  Summing these difference
images, we find a very clear stellar profile located $1.\hskip-2pt'' 1
\pm 0.\hskip-2pt'' 1$ south of the source.  Hence, this bump is due to
genuine stellar variability, which was close enough to contaminate our
$2.\hskip-2pt'' 3$ superpixels, but is clearly distinguishable from
the event source.  We eliminate these contaminated December points
from the analysis.

\subsection{PA-99-N1 as a Microlensing Event}

	Making use of the identification of the {\it HST} star as the
source of the microlensing event, we fit the data to a full
(non-degenerate) microlensing curve and so evaluate the Einstein
crossing time, $\te=10.4\,$days.

\begin{table*}
\begin{center}
\begin{tabular}{ccccccccc}
\hline
\multicolumn{9}{c}{Fit Parameters} \\
\hline
$N$ & $\chi^2$ & $\tp$ & $u_0$ & $\te$  
& $F_{s,r'}$ & $F_{s,g'}$ & $F_{b,r'}$ & $F_{b,g'}$ \\
&& (days) && (days) & (ADU/s) & (ADU/s) & (ADU/s) & (ADU/s) \\\hline
$97$ & $154$&$13.87$ & $0.056$ & $10.4$ & $1.02$ & $0.28$ & $397.65$ & $209.40$
\\
& $\pm$   &   $ 0.04$ & $0.009$ & $1.5$ & $0.17$ & $0.05$ & $  0.18$ & $  0.10$
\\\hline
\end{tabular}
\end{center}
\caption{Fit parameters for PA-99-N1.
$N$: number of points, $\tp$: time of the peak,
$u_0$: impact parameter, $\te$: Einstein timescale,
$F_s$ and $F_b$: source and background fluxes
in two bands.  The maximum magnification is $\Ap\sim u_0^{-1}\sim 18$}.
\label{table}
\end{table*}

By permitting the measurement of $\te$,
\begin{equation}
\te={\thetae\over\mu_\rel},\qquad
\thetae=\sqrt{\pi_\rel \left({4 GM/ c^2 \over \rm AU}\right)},
\label{eq:thetae}
\end{equation}
the identification of the source reduces but does not eliminate the
degeneracy in the lens parameters.  The mass $M$ remains entangled
with $\pi_\rel$ and $\mu_\rel$, the lens-source relative parallax and
proper motion.  If the lens transits the source, one can measure
$\mu_\rel$ and so further break the parameter degeneracy (Gould 1994;
Alcock et al.\ 1997).  We are not able to measure $\mu_\rel$ for
PA-99-N1, but are able to strongly constrain it
from the lack of finite source effects that would be induced by
a transit.

	Let $\rho_*=\theta_*/\thetae$, where $\theta_*$ is the angular
radius of the source.  We find that if we fit the lightcurve with
$\rho_*$ fixed at any value $\rho_*\leq 0.1$, then the other
parameters all assume the same values as in Table 1.  However, as
$\rho_*$ is increased further, $\te$ declines, so that the parameter
combination $\rho_* \te$ saturates at $1.0\,$day, and then for
$\rho_*\ga 0.12$, $\chi^2$ rises dramatically.  From this we derive
the constraint $\rho_*\te\leq 1.0\,$day, implying,
\begin{equation}
\mu_{\rm rel}  = {\thetae \over \te} = {\theta_*\over \rho_* \te}
\geq {\theta_*\over 1.0\,\rm day}.
\label{eq:mueqn}
\end{equation}
Using the empirical surface-brightness/color relation of van Belle
(1999), the color-color transformation of Bessell \& Brett (1988), and
an estimated total extinction $A_V=0.4\pm 0.2$, we find $\theta_\star
= 0.30 \pm 0.04\, \mu{\rm as}$, where the uncertainty is dominated by
the 11\% intrinsic scatter about the van Belle (1999) relation.  The
constraint (\ref{eq:mueqn}) then yields
\begin{equation}
v_{\rm rel} = \mu_\rel D_{\rm l} > 400~{\rm km\,s}^{-1} \left( {\rho_* \te
 \over 1.0\ {\rm day}} \right)^{-1} \left( {\theta_\star \over 0.3~
\mu{\rm as}} \right) \left( {D_{\rm l} \over 770~{\rm kpc}} \right),
\label{eq:finite}
\end{equation}
where $v_\rel$ is the transverse speed of the lens relative to the
observer-source line of sight, and $D_l$ is the distance to the lens.
While the constraint (\ref{eq:finite}) is unimportant for Galactic lenses
(where $D_{\rm l} \sim 10-30$ kpc), it strongly limits the allowed
range of $\mu_\rel$ for M31 lenses (since high values are
exponentially suppressed).  Hence $\mu_\rel$ is crudely measured,
which implies, via equation (\ref{eq:thetae}), a constrained relation
between $M$ and $\pi_\rel$.  Thus, when the proper-motion constraint
(\ref{eq:finite}) is incorporated into the Monte Carlo (see below) it
indirectly constrains the mass to a much narrower range than would be
allowed without it.  Note that our finite source fits incorporate a
linear limb-darkening parameter of 0.78 in $R$ band, which is
appropriate for a $V-I=2.1$ (M0-M1, 3630 K) star (Manduca, Bell \&
Gustafsson 1977; Johnson 1966).

\section{Discussion}

The location of PA-99-N1, at nearly $8'$ from the center of M31, is
interesting because the majority of stellar lenses are expected to
reside within $5'$ of the M31 center (e.g., Kerins et al.\ 2001).  The
significance of the candidate is assessed by performing Monte Carlo
simulations of events with the same source magnitude, Einstein
timescale and projected position as PA-99-N1, as well as taking
account of the lower limit on $\mu_\rel$ in
equation~({\ref{eq:finite}).  We use the actual sampling and exposure
times for the 1999 season.  We model the halos of both galaxies with
cored near-isothermal spheres, taking the mass of M31 as twice the
mass of the Galaxy and assuming a core radius of 5 kpc for both
galaxies.  The M31 bulge follows Kent's (1989) axisymmetric model
while the disk has a sech-squared profile (see Kerins et al.\ 2001 for
details).  We model the disk stellar lens masses based on the Galactic
disk mass function (MF) of Gould, Bahcall, \& Flynn (1997) corrected
for binaries and extended down to $0.01\,M_\odot$, and the bulge
stellar masses using the Zoccali et al.\ (2000) Galactic bulge MF,
similarly corrected and extended.

The thick solid line in Figure \ref{fig:lim} (a) shows the relative
probability $P$ that PA-99-N1 is due to a Macho of mass $M$, assuming
full Macho halos. The curve is normalized with respect to the
probability that the lens is a star, $P_\star$. The thick dashed
(dotted) curve gives the contribution to $P(M)$ from M31 (Milky Way)
Machos.  In addition to this, we have plotted the M31 distribution
before applying the proper motion cut of equation~(\ref{eq:finite})
(thin dashed line). The black horizontal line (`d/b') gives the
fractional contribution of the disk lens and bulge source
configuration to $P_\star$, whilst the gray line (`b/b') shows the
contribution of bulge-bulge lensing. The width of these lines reflects
the dispersion in the logarithm of the stellar lens mass.

Clearly, if the halos are full of Machos, the most probable
interpretation is that the lens is a Macho with mass $\sim 0.03
\msun$. This is about five times more likely than PA-99-N1 being due
to a stellar lens.  The lens is about equally likely to lie in the Milky
Way or M31 halo, which is quite unexpected given that the line of
sight passes only $2$ kpc from the center of M31, but $8$ kpc from the
Galactic center.  Normally this would cause the M31 probability to be
much higher (as shown by thin dashed line).  However, the
$\mu_\rel$ constraint (\ref{eq:finite}) severely suppresses the M31
distribution at low masses, while leaving the Milky Way distribution
virtually unaffected. This is because the average Milky Way Macho has
a relative proper motion of $\sim 10~\mu$as/day, whilst
for the typical M31 Macho it is $\sim 0.2~\mu$as/day.  The
constraint also has the effect of displacing the peak of the M31
distribution towards higher mass. 

Assuming a logarithmic prior in the Macho mass gives $M =
6.4^{+21}_{-4.9}\times 10^{-2}\msun$ for an M31 Macho and
$2.1^{+7.0}_{-1.6} \times 10^{-2}\msun$ for a Milky Way Macho.  The
M31 and Milky Way distributions are both quite broad.  As they are
about equal in amplitude and are displaced from one another by nearly
a decade in mass, their combined distribution is even broader, with a
FWHM extending from the planetary to the stellar mass range.

If the lens is a star, then the overall mass probability distribution
is given by the thick solid line in Figure \ref{fig:lim} (b).  The
thick dashed line (`d/b') shows the contribution from disk-bulge
lensing whilst the thick dotted line (`b/b') shows the distribution
for bulge-bulge lensing. The contribution of lensing involving disk
sources is negligible. In the absence of the relative proper motion
cut, the disk and bulge distributions would be given by the thin
dashed and dotted lines respectively.  Including these cuts results in
a most probable mass around $0.2 \msun$, making a rather plausible
hypothesis that the lens is a low mass star. Logarithmic averaging of
the solid line gives $M = 0.27^{+0.21}_{-0.12}\msun$, i.e., only a
factor 1.8 ($1\,\sigma$) uncertainty.

	We see that, if the lens is an M31 star, it most likely lies
in the disk.  This is a direct consequence of the fact that the source
lies at the edge of the bulge, while the line of sight passes through
the near side of the disk.  Comparing Figures \ref{fig:lim}a and
\ref{fig:lim}b, it is clear that the mass of the lens is much better
constrained if it lies in the M31 disk than the M31 halo.  This is
because for M31 lenses, $M\propto (\mu_\rel t_{\rm E} D_{\rm ls})^2$
where $D_{\rm ls}$ is the distance from the lens to the source (see
eq.\ [\ref{eq:thetae}]).  For both populations, the product $\mu_\rel
t_{\rm E}$ is reasonably well constrained, but for halo lenses,
$D_{\rm ls}$ can take on a very broad range of values, while for disk
lenses, the geometry implies $D_{\rm ls}=4.0\pm 1.8\,$kpc.  Since
$t_{\rm E}$ is measured and $\mu_\rel$ is constrained by equation
(\ref{eq:finite}), $M$ is also constrained. This means that we only
detect events at the location and timescale of PA-99-N1 if the stellar
mass $M > 0.1 \msun (4~{\rm kpc}/ D_{\rm ls} )$.

The relative rate of Machos to stars is subject to a number of
modeling uncertainties.  First, we have assumed that the M31 halo is
twice as massive as that of the Milky Way. The most recent mass
estimates of the M31 halo derived from the kinematics of the satellite
galaxies suggest that it may be only roughly as massive as the Milky
Way (Evans et al.\ 2000). Second, it is sensitive to the choice of the
core radius for M31's halo, with larger core radii giving a reduced
M31 Macho rate at the location of PA-99-N1.  Third, we have assumed
that the M31 bulge is axisymmetric, but the twisting of the optical
isophotes (e.g., Walterbos \& Kennicutt 1987) is evidence for the
presence of a bar. Inspection of their Figure 5 shows that the
twisting is away from the location of PA-99-N1. We therefore expect a
bar model with the same overall mass as our axisymmetric model to have
a lower surface density at the position of the event and thus a lower
stellar lensing rate there.

\section{Conclusions}

We have reported the discovery of PA-99-N1, a high signal-to-noise,
short duration event which is consistent with the microlensing
hypothesis. Almost certainly, the source star has been identified on
archival {\it HST} frames, from which the Einstein crossing time of
$10.4$ days has been determined. We have argued -- from the stability
of the lightcurve, the achromaticity of the flux excess, the
excellence of the fit, and the consistency of the color of the event at
maximum with the color of the {\it HST} source -- that by far the most
natural explanation of this event is microlensing. The lens is most
likely to be a Macho if the halo fraction is above $20 \%$. However,
it is also plausible that the lens is a disk star with a mass of
$\sim 0.2 \msun$.

\bigskip
\noindent
{\bf Acknowledgments}: YLD, EJK and SJS are supported by PPARC
postdoctoral fellowships and EL by a CNPq postdoctoral fellowship. NWE
acknowledges financial support from the Royal Society.  Work by AG was
supported in part by a grant from Le Minist\`ere de l'Education
Nationale de la Recherche et de la Technologie and in part by grant
AST 97-27520 from the NSF.

\begin{figure}
\epsfysize=15cm \centerline{\epsfbox{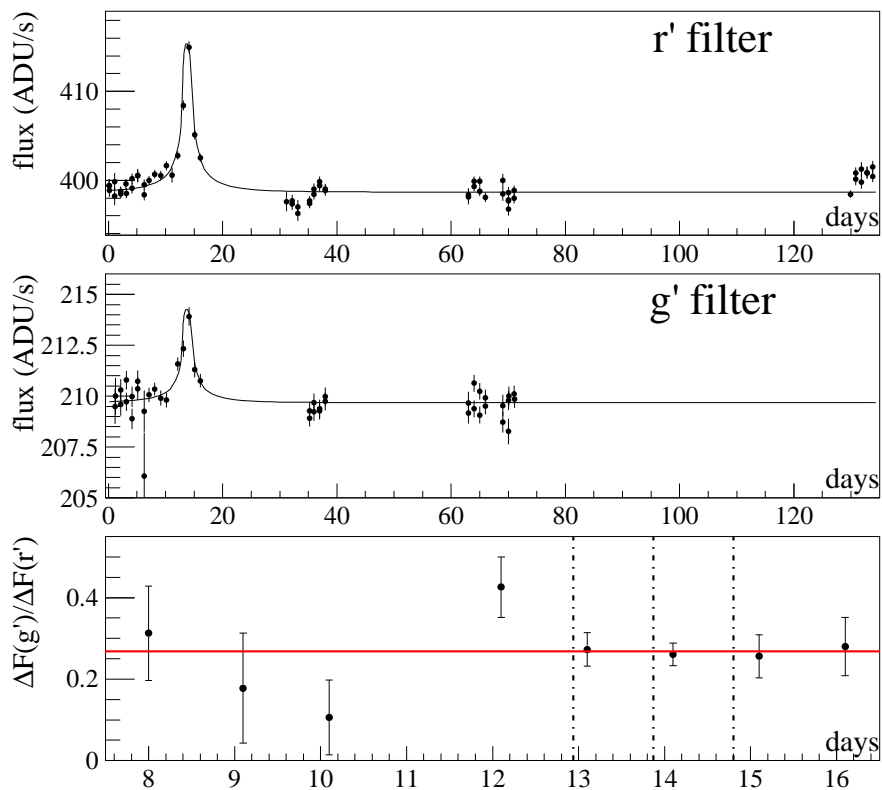}}
\caption{ Panels (a) and (b) show the flux in $r'$ and $g'$ against
time in days. Panel (c) is a zoom centred on the event that shows the
variation of the ratio of the flux change in the two passbands $\Delta
F_{g'}/\Delta F_{r'}$ with time. The vertical lines are centred on
$t_0$ and are separated by 0.9 days, i.e., half the full width at
half-maximum.  The days correspond to $J - 2451392.5$ where $J$ is the
Julian date. }
\label{fig:lightcurves}
\end{figure}
\begin{figure}
\begin{center}
\psfig{file=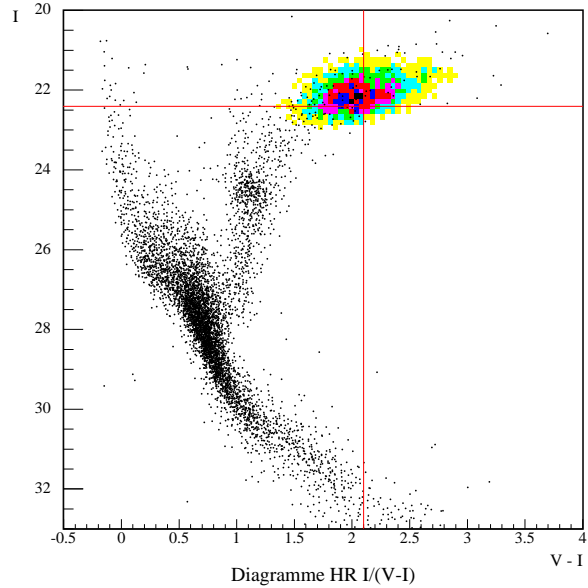,width=0.5\textwidth}\\
\end{center}
\caption{Superposition of the Hipparcos (points) and {\it HST} (shaded
area) color-magnitude diagrams.  The Hipparcos stars have been moved
to the position of M31 [assuming $(m-M)_0=24.43$ and Galactic
foreground extinction $A_V=0.24$] . The intersection of the horizontal
and vertical lines shows the position of the HST object.}
\label{hst1}
\end{figure}

\begin{figure}
\begin{center}
\psfig{file=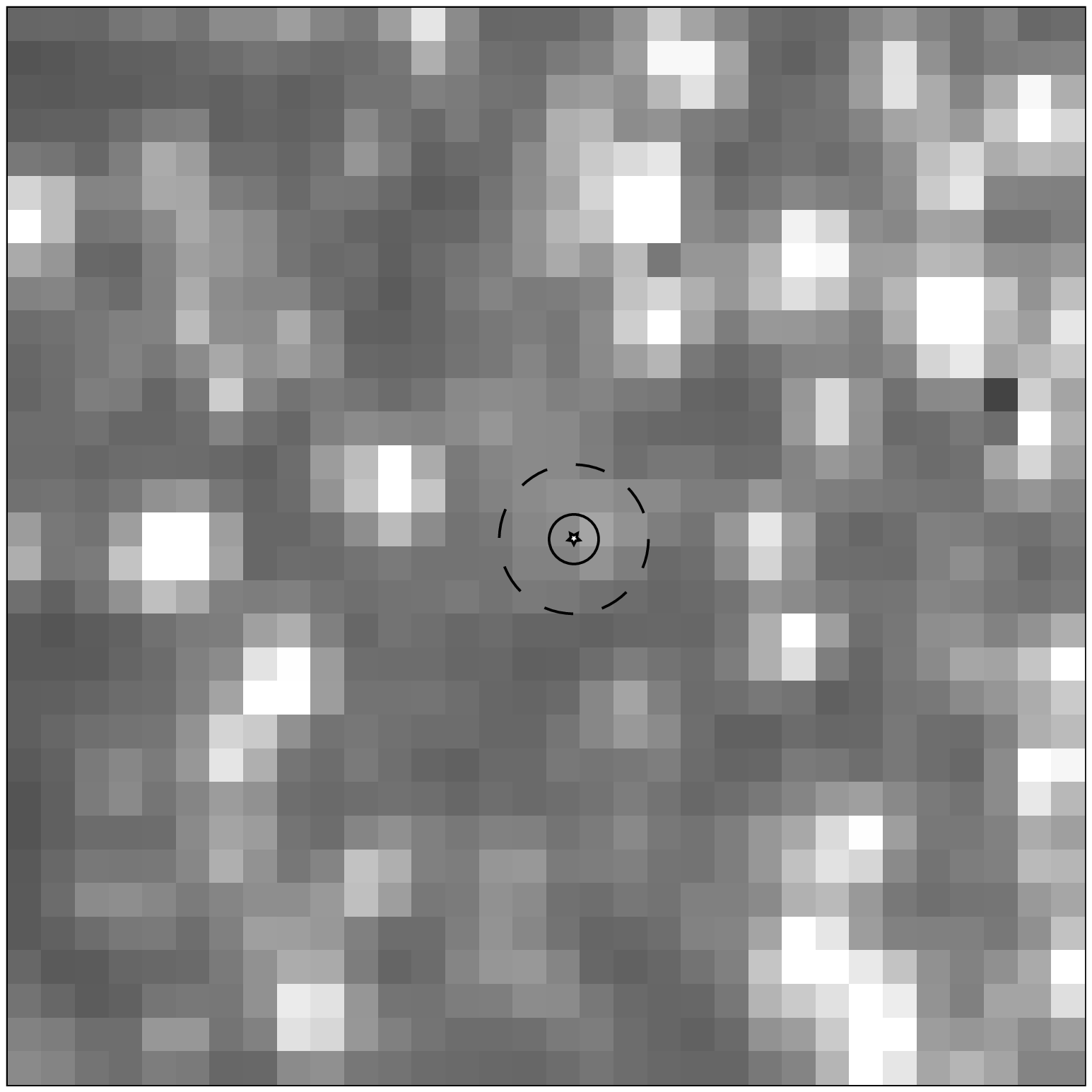,clip,width=0.3\textwidth}\\
\end{center}
\caption{An image of the HST field showing the $1\, \sigma$ and $3\,
\sigma$ error circles around the position of the event. There is a
resolved source close to the $1\,\sigma$ circle.}
\label{hststar}
\end{figure}

\begin{figure}
\begin{center}
\epsfig{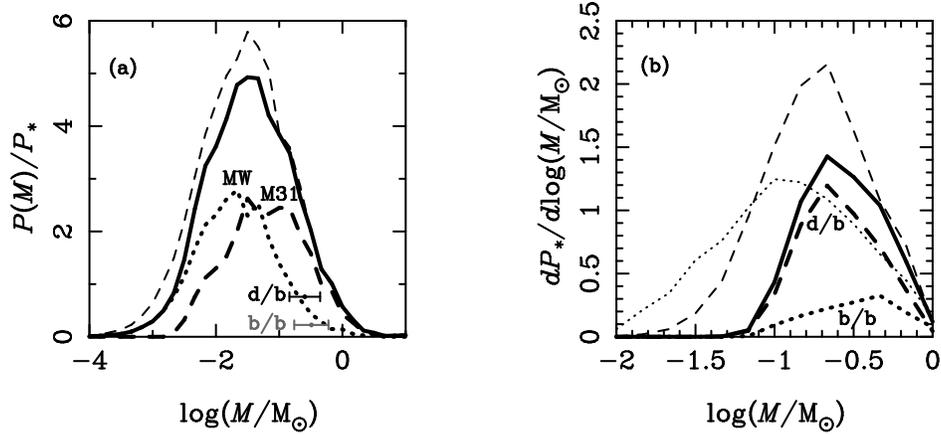}\\
\end{center}
\caption{Panel (a) shows, for the case where PA-99-N1 is a Macho, the
probability $P$ that the lens has mass $M$ (thick solid line),
normalized to the probability $P_\star$ that PA-99-N1 is a stellar
lens. The contribution to this curve from M31 (Milky Way) Machos is
shown by the thick dashed (dotted) line. The thin dashed line shows
$P(M)$ for M31 Machos before the relative proper motion cut (eq.\
\ref{eq:finite}).  The horizontal bars show the contribution to
$P_\star$ from bulge-bulge lensing (`b/b') and disk-bulge lensing
(`d/b'), whilst their width indicates the logarithmic dispersion in
the stellar lens mass. Panel (b) shows, for the case where PA-99-N1 is
a stellar lens, the overall stellar probability density as a function
of lens mass (solid line), including the contributions from
bulge-bulge lensing (thick dotted line) and from disk-bulge lensing
(thick dashed line). The thin dotted and dashed lines show the
relative distributions of bulge and disk lenses before the proper
motion cut.}
\label{fig:lim}
\end{figure}


\begin{thebibliography}{}

\bibitem[Alcock et al.]{alcocka} Alcock C., et al., 1997, \apj, 491, 436

\bibitem[Alcock et al.]{alcockb} Alcock C., et al., 2000, \apj, 542, 281

\bibitem[Allen 1999]{allen} Allen C., 1999, Astrophysical Quantities,
4th edition (ed. A.N. Cox)

\bibitem[Ansari et al.\ 1997]{ans97}
Ansari R., et al., 1997, \aap, 324, 843

\bibitem[Ansari et al.\ 1999]{ans99}
Ansari R., et al., 1999, \aap, 344, L49

\bibitem[Baillon et al.\ 1993]{bai93}
Baillon P., Bouquet A., Giraud-H{\'e}raud Y., Kaplan J., 1993, \aap,
277, 1

\bibitem[Bessell \& Brett]{bb} Bessell M.S., Brett J.M., 1988,
\pasp, 100, 1134

\bibitem[Capaccioli et al.\ 1989]{cap} Capaccioli M., Della Valle M.,
D'Onofrio M., Rosin L., 1989, \aj, 97, 1622

\bibitem[Crotts 1992]{cro92}
Crotts A.P.S., 1992, \apj, 399, L43

\bibitem[Crotts \& Tomaney 1997]{cro97}
Crotts A.P.S., Tomaney A.B., 1997, \apj, 473, L87

\bibitem[Demarque et al.\ 1996]{dcgppy}
Demarque, P., Chaboyer, B., Guenther, D., Pinsonneault, M, Pinsonneault,
L., \& Yi, S.\ 1996, Yale Isochrones 1996, at http://achee.srl.caltech.edu/

\bibitem[Evans et al.\ 2000]{evans} Evans N.W., Wilkinson M.I., 
Guhathakurta P., Grebel E.K., Vogt S.S., 2000, \apj, 540, L9

\bibitem[Gould 1994]{andyga} Gould A., 1994, \apj, 421, L71

\bibitem[Gould 1996]{andygb} Gould A., 1996, \apj, 470, 201

\bibitem[GBF 1997]{gbf} Gould A., Bahcall, J.N., \& Flynn, C.\
1997, \apj, 482, 913

\bibitem[Haiman et al.\ 1994]{haiman} Haiman Z., Magnier E.A., Lewin
W.H.G., Lester R.R., van Paradijs J., Hasinger G., Pietsch W., 
Supper R., Tr\"umper J., 1994, \aap, 286, 725


\bibitem[Johnson 1966]{johnson} Johnson H.K., 1966, \araa,4, 193

\bibitem[Kent 1989]{kent} Kent S.M., 1989, \aj, 97, 1614

\bibitem[Kerins et al.\ 2001]{eamonn}
Kerins E.J., Carr B.J., Evans N.W., Hewett P.C., Lastennet E.,
Le~Du Y., Melchior A., Smartt S., Valls-Gabaud D., 2001,
\mnras, in press (astro-ph/0002256)

\bibitem[Lasserre et al 2000]{lasserre} Lasserre T. et al., 2000,
\aap, 355, L39

\bibitem[Le~Du 2000]{ledu00}
Le~Du Y., 2000, Ph.D thesis, University Paris VI, Coll\`ege de France

\bibitem[Magnier et al.\ 1992]{magone} Magnier E.A., Lewin W.H.G., 
van Paradijs J., Hasinger G., Jain A., Pietsch W., Tr\"umper J., 1992, 
\aaps, 96, 379

\bibitem[Magnier et al.\ 1993]{magtwo} Magnier E.A., Lewin W.H.G.,
van Paradijs J., Hasinger G., Pietsch W., Tr\"umper J., 1993,
\aap, 272, 695 

\bibitem[Manduca et al.\ 1977]{man} Manduca A., Bell R.A., Gustafsson B.,
1977, \aap, 61, 809

\bibitem[Paczy\'nski]{Paczynski86} Paczy\'nski B., 1986, \aj, 304, 1 

\bibitem[Stetson 1987]{stetson} Stetson P. B., 1987, \pasp, 99, 191

\bibitem[Strauman et al.\ 1998]{naples98} Strauman N., Jetzer Ph.,
Kaplan J., 1998, ``Topics on gravitational lensing'', Napoli Series in
Physics and Astrophysics, 1, 102.

\bibitem[van Belle]{vb} van Belle G., 1999, PASP, 111, 1515

\bibitem[Walterbos \& Kennicutt]{wk} Walterbos R.A.M., Kennicutt
R.C., 1987, \aaps, 69, 311


\bibitem[Yi et al.]{ydo} Yi, S., Demarque, O., \& 
Oemler, A., Jr.\ 1997, \apj, 486, 201

\bibitem[Zheng et al.]{zheng} Zheng, Z., Flynn, C., Gould, A., 
Bahcall, J.N., \& Salim, S.\ 2000, \apj, submitted

\bibitem[Zoccali et al.\ 2000]{zoc} Zoccali M., Cassisi S., Frogel J.A.,
Gould A., Ortolani S., Renzini A., Rich  R.M., Stephens A.W., 2000,
\apj, 530, 418


\end{thebibliography}
\end{document}